\documentclass[preprint,12pt, nofootinbib]{elsarticle}

\usepackage{amssymb}
\usepackage{amsmath}
\usepackage{hyperref}
\usepackage{color}
\usepackage{setspace}
\usepackage{pdfsync}

\def\sideremark#1{\ifvmode\leavevmode\fi\vadjust{\vbox to0pt{\vss
 \hbox to 0pt{\hskip\hsize\hskip1em
 \vbox{\hsize3cm\tiny\raggedright\pretolerance10000
 \noindent #1\hfill}\hss}\vbox to8pt{\vfil}\vss}}}%

\journal{${}$}

\begin{document}


\begin{frontmatter}



\title{Massive Gravity Acausality Redux}


\author[everywhere,nowhere]{S. Deser\fnref{Stanley}}
\author[here]{K. Izumi\fnref{Keisuke}}
\author[here,there]{Y. C. Ong\fnref{Yen Chin}}
\author[somewhere]{A.~Waldron\fnref{Andrew}}

\fntext[Stanley]{\href{mailto:deser@brandeis.edu}{deser@brandeis.edu}}
\fntext[Keisuke]{\href{mailto:izumi@phys.ntu.edu.tw}{izumi@phys.ntu.edu.tw}}
\fntext[Yen Chin]{\href{mailto:ongyenchin@member.ams.org}{ongyenchin@member.ams.org}}
\fntext[Andrew]{\href{mailto:wally@math.ucdavis.edu}{wally@math.ucdavis.edu}}

\address[everywhere]{Lauritsen Lab, Caltech, Pasadena CA 91125, USA.}

\address[nowhere]{Physics Department, Brandeis University, Waltham, MA 02454, USA.}

\vspace{20mm}

\address[here]{Leung Center for Cosmology and Particle Astrophysics,\\ National Taiwan University, Taipei 10617, Taiwan.}

\address[there]{Graduate Institute of Astrophysics, National Taiwan University, Taipei 10617, Taiwan.
}

\address[somewhere]{Department of Mathematics, University of California, Davis, CA 95616, USA.
}

\begin{abstract}
Massive gravity (mGR) is a~$5(=\!2s\!+\!1)$ degree of freedom, finite range extension of GR.
However, amongst other problems, it is plagued by superluminal propagation, first uncovered via a second order shock analysis. First order mGR shock structures have also been studied, but the existence of superluminal propagation in that context was left open. We present here a concordance of these methods, by an explicit (first order) characteristic matrix computation, which confirms mGR's superluminal propagation  as well as acausality.
\end{abstract}




\end{frontmatter}
{\flushright BRX-TH668, CALT 68-2942}



\section{Introduction}
A natural physical question is whether gravity is necessarily inf\-inite ran\-ge---like its non-abelian Yang--Mills (YM) counterpart---or whether ``nearby'', massive, 
extensions are also permitted, at least as effective theories within a certain domain of validity. This question was first studied at linearized level almost 80 years ago by Fierz and Pauli (FP)~\cite{FP}, who constructed a massive spin~$s=2$ model with the required~$2s+1=5$ degrees of freedom~(DoF). Even this was nontrivial, as the ``natural'' DoF count would be six--the number of components of the symmetric 3-tensor~$h_{ij}$ governing the kinetic, linearized Einstein, action. Indeed, (up to field redefinitions) only one mass combination,~$m^2 (h_{\mu\nu}\bar g^{\nu\rho}h_{\rho\sigma}\bar g^{\sigma\mu}  - h_{\mu\nu}\bar g^{\mu\nu}h_{\rho\sigma}\bar g^{\rho\sigma})$, accomplishes this so long as the fiducial metric~$\bar g_{\mu\nu}$ is Einstein ($G_{\mu\nu}(\bar g)\propto \bar g_{\mu\nu}$)~\cite{BB}. The (observationally necessary) extension to the nonlinear domain, with the full scalar curvature~$R(g_{\mu\nu})$ kinetic term and mass terms built from an arbitrary (diffeomorphism invariant) combination of the dynamical metric~$g_{\mu\nu}$ and the fixed (but now potentially arbitrary) background~$\bar g_{\mu\nu}$, proved more elusive. 

Further developments began about halfway since the time of FP, but almost immediately ground to a halt because it was shown that, for generic mass terms, a sixth, ghost, excitation necessarily develops beyond linear, FP, order~\cite{BD}. This was catastrophic because this ghost arises \emph{within} the effective theory's supposed domain of validity, reducing it to nil. 
It took the subsequent four decades to discover that exactly three mass terms evade this no-go result.
One of these was discovered in~\cite{WZ} based upon the bimetric model of~\cite{Salam}. 
Much later, that mass term and two others were uncovered
in mGR's decoupling limit~\cite{dRGT}. Absence of the ``bulk'' ghost mode was finally proven in~\cite{HR}.
Predictably, it was time for the next blow to strike: The very mass terms that avoided the ghost replaced that woe with superluminal--tachyonic modes, discovered by analyzing second order shocks~\cite{DW}. This result was perhaps not surprising\footnote{It also follows a similar pattern of massive higher spin inconsistencies when these models interact with background fields; see old results for~$s=3/2,2$ in both E/M and GR backgrounds~\cite{bbd}.} since superluminal behavior
had already been uncovered in the model's St\"uckelberg sector and decoupling limit~\cite{Gr} as well as in a spherically symmetric analysis on Friedmann-Lema\^itre-Robertson-Walker~(FLRW) backgrounds~\cite{Chien-I}. Concordantly, unstable cosmological solutions were discovered~\cite{MU}
(similar pathologies also arise in other nonlinear gravity models, such as~$f(T)$~\cite{OINC} and Poincar\'e gauge gravity~\cite{Nester8}). Moreover, mGR also seems not to allow static black hole solutions~\cite{BF}.

The characteristics of mGR were subsequently studied in~\cite{IO} in a certain first order formulation where a (generically) maximal rank characteristic matrix was found. However, a study of  zeros of that matrix and thus superluminality was postponed in that work, which focused on the relationship between first order shocks and the second order shocks of~\cite{DW}. In this work, we exhibit further superluminal behavior in the first order setting and clarify the relation between the various superluminal modes and acausality. We also give a  compact computation and formula for the (pathological)~mGR characteristic matrix by employing vierbeine and spin connections. A toy scalar field example is given in the discussion, which further illuminates our findings. The power of the characteristic  method~\cite{bbd} is that there is no need to wait the thirty odd years it took 
for G\"odel to discover closed timelike curves in GR, but rather acausality can be detected without directly solving the mGR field equations\footnote{Actually, in~\cite{Chien-I}, solutions with  infinitely rapid propagation---in open FLRW backgrounds---were explicitly given; 
 these are likely to include examples of acausal structures, though their energy scale is as yet unclear.}. 
Moreover the causal inconsistencies we find are local, as opposed to the  non-local G\"odel type acausal anomalies of GR.
Our conclusion is that mGR is~\emph{unphysical}, leaving GR on its isolated consistency pedestal.

\section{Massive  Gravity}

The model's field equation is
\begin{equation}\label{EOM0}
G_{\mu\nu}(g)=\tau_{\mu\nu}(f,g):=\Lambda g_{\mu\nu}-m^2\Big(f_{\mu\nu}-g_{\mu\nu} f\Big)\, ,
\end{equation}
where the metric~$g_{\mu\nu}$ is dynamical and~$G_{\mu\nu}(g)$ is its Einstein tensor. The rank two tensor 
$$
f_{\mu\nu}:=f_{\mu}{}^m e_{\nu m}
$$
is built from the vierbein~$e_\mu{}^m$ of the dynamical metric~$g_{\mu\nu}$ and a non-dynamical vierbein~$f_\mu{}^m$
of a non-dynamical {\it background/fiducial} metric~$\bar g_{\mu\nu}$. All index manipulations will be performed using 
the dynamical metric and vierbein, in particular~$f:=f_\mu{}^m e^\mu{}_m$. The inverse background vierbein is denoted by~$\ell^\mu{}_m$.

Of the three permitted bulk ghost-free mass terms, we  focus on the above, simplest, possibility (linear in the fiducial vierbein); 
of the other two, one is known to have tachyonic behavior as well~\cite{DSW}, while the last is---formally---open because its covariant constraint form, if any, is as yet unknown~\cite{DMZ}. 

The parameter~$m$ is the FP mass when the theory is linearized 
around an Einstein background~$\bar g_{\mu\nu}$  with cosmological constant~$\bar\Lambda$. Requiring a good linearization  (without constant terms in the linear equations of motion) demands the further parameter condition~$\Lambda-\bar \Lambda +3m^2=0$ (in particular flat backgrounds are achieved by tuning the {\it parameter} $\Lambda=-3m^2$).
Also, we have denoted~$f:=f^\mu_{~\mu}$ and, as a consequence of Eq.~(\ref{EOM0}), the vierbein obeys the symmetry constraint 
\begin{equation}\label{symmetry}
f_{[\mu}{}^m e_{\nu]m}=0\, .
\end{equation}

\section{First Order Formulation}

To perform a first order shock and characteristic surface analysis  we first write the system
in a first order formulation in the usual way. The dynamical metric 
$g_{\mu\nu}$ is replaced by the vierbein~$e_{\mu}{}^m$ (with
$
g_{\mu\nu}=e_\mu{}^m\eta_{mn}e_\nu{}^n
$),
and an off-shell spin connection~$\omega_{\mu}{}^m{}_n$ determined by the torsion-free condition built into the ``Palatini" first order action,
\begin{equation}\label{Tfree}
\partial_{[\mu} e_{\nu]}{}^m+\omega_{[\mu|}{}^m{}_n e_{|\nu]}{}^n  =0\, .
\end{equation}
The standard Bianchi identities for the Riemann tensor then become 
 first order integrability conditions
\begin{equation}\label{Bianchi}
R_{\mu\nu\rho\sigma}(e,\omega)-R_{\rho\sigma\mu\nu}(e,\omega)=0=R_{[\mu\nu\rho]\sigma}(\omega,e)\, .
\end{equation}
Note that there is no need to impose the condition ~$\nabla_{[\mu}R_{\nu\rho]\sigma\kappa}=0$ because it holds identically 
for any~$\omega$. The field equations imply that the Einstein tensor obeys~$G(e,\omega)_{\mu\nu}=G(e,\omega)_{\nu\mu}$ and, in turn, the symmetry constraint~(\ref{symmetry}). The latter's curl gives a further integrability condition
\begin{equation}\label{curl}
f_{[\mu}{}^\sigma K_{\nu\rho]\sigma}=0\, 
\end{equation}
where the contorsion,
$$
K_{\mu}{}^m{}_n:=\omega_\mu{}^m{}_n-\omega(f)_\mu{}^m{}_n\, ,
$$
measures the failure of parallelograms of one (torsion-free) connection to close with respect to the other
and will play a crucial role in further developments.

Going beyond kinematics, dynamics are generated by the first order evolution equation
\begin{equation}\label{EOM1}
G_{\mu\nu}(e,\omega)-\Lambda g_{\mu\nu}+
m^2 \big(f_{\mu\nu}-g_{\mu\nu} f\big)=0\, ,
\end{equation}
where~$G(e,\omega)$ is obtained from the Riemann tensor~$R(\omega)=d\omega+\omega\wedge\omega$ in the usual way.
 
So far the choice of couplings~$\tau_{\mu\nu}$  has not been invoked. The covariant vector and scalar constraints (whose existence was 
verified in~\cite{DMZ}) responsible for the ultimate ghost free,~$5=2s+1$,~$s=2$ DoF count, depend in an essential way on this choice\footnote{To be precise, a vector constraint exists for {\it any} algebraic coupling $\tau_{\mu\nu}$, but the condition it imposes on fields is $\tau$-dependent. The very existence of a scalar constraint hinges on the exact choice of~$\tau$. }. They have been 
calculated explicitly in~\cite{DW} and read
\begin{eqnarray}
0&=&\nabla^{\mu}[{G}_{\mu\nu}-\tau_{\mu\nu}]\ =\ m^2 \, e^\mu{}_m K_{\mu}{}^m{}_n e_\nu{}^n
 =:m^2 K_\nu\, ,
\label{vector}\\[4mm]
0 &=&\frac{1}{m^2}\, \nabla_\rho \big(\ell^{\rho\nu}\nabla^{\mu}[{G}_{\mu\nu}-\tau_{\mu\nu}])
+\frac{1}2\, g^{\mu\nu}
\,[{G}_{\mu\nu}-\tau_{\mu\nu}]\label{scalar}\\[1mm]
&=&
-\, \frac{3m^2}2 \, f
-\frac{1}{2}\,  \big[ e^\mu{}_ne^\nu{}_m \bar R_{\mu\nu}{}^{mn}+4\Lambda\big]
+\frac{1}2 \, \big[K_{\mu\nu\rho}K^{\nu\rho\mu}+K_\mu K^\mu\big]
\, . \nonumber
\end{eqnarray}
Note that the term~$K_\mu K^\mu$ in the scalar constraint can be dropped since it is the square of the vector one~(\ref{vector}).

\section{Shocks}

We investigate first order shocks by positing 
$$
[\partial_\alpha e_\mu{}^m]_\Sigma=\xi_\alpha{\cal E}_\mu{}^m\, ,\qquad 
[\partial_\alpha \omega_\mu{}^m{}_n]_\Sigma=\xi_\alpha \Omega_\mu{}^m{}_n\, .
$$
Since we wish to study superluminal propagation, we take the normal~$\xi$ to be timelike:~$\xi^\mu g_{\mu\nu} \xi^\nu=-1$. 
For compactness of notation, we denote the contraction of~$\xi$ on an index of any tensor by an ``$o$'', so~$\xi. V:=V_o$, where we use lower dot to denote tensor contraction, to avoid confusion with the usual vector dot product.  
Also, the operator~$\top_\mu^\nu := \delta_\mu^\nu+\xi_\mu\xi^\nu$ is a projector; we will denote  its action
on tensors by latin indices, for example
$$
V_i:=\top_i^\nu V_\nu\, \Rightarrow V_\mu V^\mu = V_i V^i - V_o V_o\, .
$$

We split our shock analysis into two parts: First, we deal with the consequences of the ``kinematical''
equations, namely Eqs.~(\ref{Tfree}--\ref{Bianchi}), and then turn to the dynamical equation, Eq.~(\ref{EOM1})
and its constraints given by Eqs.~(\ref{symmetry},\ref{curl},\ref{vector},\ref{scalar}). These will give algebraic conditions
on the shock profiles~${\cal E}_\mu{}^m$ and~$\Omega_\mu{}^m{}_n$; there would be causal consistency only  if these conditions forced all  shock profiles to vanish. 

Firstly, we observe that the discontinuity in the torsion-free condition~(\ref{Tfree}) implies
$$\xi_{[\mu} {\cal E}_{\nu]\rho}=0\, .$$
Multiplying by~$\xi^\mu$  we find 
$$
{\cal E}_{\mu\nu} =-\xi_\mu {\cal E}_{o\nu}\, .
$$
Thus ~${\cal E}_{ij}=0={\cal E}_{io}$, so  of the  vierbein shock profiles, only ~${\cal E}_{oj}$ and~${\cal E}_{oo}$ remain.

The discontinuities of the Bianchi identities~(\ref{Bianchi}) are obtained from that of Riemann curvature tensor: 
$$[R_{\mu\nu}{}^m{}_n]_\Sigma=\xi_\mu \Omega_\nu{}^m{}_n -\xi_\nu \Omega_\mu{}^m{}_n\, .$$
Hence 
$$
\xi_{[\mu}\Omega_{\nu]\rho\sigma}-\xi_{[\rho}\Omega_{\sigma]\mu\nu}=0=\xi_{[\mu}\Omega_{\nu\rho]\sigma}\, .
$$
Contracting  these with~$\xi$ yields
$$
\Omega_{\mu\nu\rho}=-\xi_\mu \Omega_{o\nu\rho} + 2 \xi_{[\nu}\Omega_{\rho]\mu o}\, ,\quad
\Omega_{[\mu\nu]o}=\xi_{[\mu|}\Omega_{oo|\nu]}\, .
$$ 
As a consequence,
$\Omega_i=-\Omega_{ooi}$ where~$\Omega_\mu:=\Omega_\nu{}^\nu{}_\mu$.

Next we consider the dynamical equation of motion~(\ref{EOM1}), whose discontinuity implies
$$
\xi_\mu \Omega_\nu + \Omega_{\mu\nu o}- g_{\mu\nu} \Omega_o=0\, .
$$
The trace of this says~$\Omega_o=0$, thus~$\Omega_{\mu\nu o}=-\xi_\mu \Omega_{oo\nu}$.
Hence we have
$$
\Omega_{\mu\nu\rho}=-\xi_{\mu}\Omega_{o\nu\rho}\, .
$$
Therefore~$\Omega_{ijk}=0=\Omega_{ijo}$, leaving just~$\Omega_{ojk}$ and~$\Omega_{ook}$ for the spin connection shock profiles.

Now we turn to the constraints. To study the symmetry constraint, (following~\cite{IO}) we define new variables for
the vierbein shock profiles
$$
{\cal F}_{\mu\nu}:= {\cal E}_{\mu\rho}f_{\nu}{}^\rho\, .
$$
Invertiblity of~$f_\mu{}^m$ implies that the variables~${\cal F}$ are in one-one correspondence with~${\cal E}$.
From the above we already know that
$
{\cal F}_{i\nu}=0
$.
In the new variables, the jump in the symmetry constraint~(\ref{symmetry}) gives
$
{\cal F}_{[\mu\nu]}=0\
$.
These two relations imply that~${\cal F}_{io}=0={\cal F}_{oi}={\cal F}_{ij}$, {\it i.e.},~${\cal F}_{\mu\nu}=\xi_\mu\xi_\nu {\cal F}_{oo}$.

At this point, only the spin connection shock profiles~$(\Omega_{ojk},\Omega_{ook})$ and~${\cal F}_{oo}$ remain. The discontinuity in the 
 vector constraint~(\ref{vector}) is easily computed
 \begin{eqnarray}
\Omega_{\rho}- {\cal E}^{\nu\mu} K_{\mu\nu\rho}=0\, .
\end{eqnarray}
This produces a relation between~$\Omega_{ook}$ and~${\cal F}_{oo}$:
\begin{eqnarray*}
\Omega_{ook}-\ell_{o}^\mu K^{\phantom \mu}_{\mu ok}{\cal F}_{oo}^{\phantom \mu}=0\, .
\end{eqnarray*}
Our characteristic analysis is now almost complete, since this relation allows us to determine~$\Omega_{ook}$, leaving
only the profiles~$\Omega_{ojk}$ and~${\cal F}_{oo}$. We  stress that up to this point all other shock profiles 
have been determined {\it algebraically} in terms of these by relations that are everywhere invertible in field space. This will 
no longer be the case for  the system of equations obeyed by~$(\Omega_{ojk},{\cal F}_{oo})$.

\section{Superluminality and Acausality}
\label{sec5}

Our shock analysis is completed by studying the discontinuities in the scalar constraint~(\ref{scalar}) 
and the curl of the symmetry constraint~(\ref{curl}):
\begin{eqnarray*} \label{super}
0&=& -\Big[\frac{3 m^2}{2}-\ell^{\mu}_o \big({\bar R}_{\mu\nu}{}^{\nu}{}_o+K_{\mu\nu\rho} K^{\nu\rho}{}_o\big)\Big]{\cal F}_{oo}  
-\Omega_{ojk}K^{jk}{}_o\, ,\\[3mm]
0&=&
f_{i}{}^k\Omega_{ojk}-f_{j}{}^k\Omega_{oik}-\big[f_i{}^\mu K_{j\nu\mu}\ell^\nu_o-f_j{}^\mu K_{i\nu\mu}\ell^\nu_o\big]{\cal F}_{oo}\, .
\end{eqnarray*}
Defining~$\tilde\Omega_i=\epsilon_{ijk}\Omega_{o}{}^{jk}$ and~$\tilde K_i=\epsilon_{ijk} K^{jk}{}_{o}$, where~$\epsilon_{ijk}:=\frac{1}{\sqrt{-g}}\, \xi^\mu \varepsilon_{\mu ijk}$ ($\varepsilon_{\mu\nu\rho\sigma}$ is the density obtained by lowering the indices of $\varepsilon^{\mu\nu\rho\sigma}$ with the dynamical metric), our characteristic determinant problem becomes 
$$
0=\begin{pmatrix}-\frac{3 m^2}{2}+\ell^{\mu}_o \big[{\bar R}_{\mu\nu}{}^{\nu}{}_o+K_{\mu\nu\rho} K^{\nu\rho}{}_o\big]& \frac 12\, \tilde K_j
\\[3mm]
[f\times K\ell]_i & f_{ij}-g_{ij} f^{(3)}
\end{pmatrix}\begin{pmatrix}{\cal F}_{oo}\\[3mm]\tilde\Omega^j\end{pmatrix}\, ,
$$
where
$[f\times K\ell]_i:=2\, \epsilon_{ijk} f^{k\mu} K^{j}{}_{\nu\mu}\ell^\nu_o$ and~$f^{(3)}:=g^{ij}f_{ij}$.
As emphasized in~\cite{Nester8}, a field-dependent characteristic matrix always forewarns of danger to consistent Cauchy propagation.
Let us analyze this in more detail.

We proceed by first assuming that the matrix~$f_{ij}-g_{ij} f^{(3)}$ is invertible, although field configurations
where even this invertibility requirement fails can occur because~$\xi_\mu$ is timelike with respect to the dynamical metric but not necessarily with respect to the background; this is responsible for the acausalities that are analyzed below.  We denote $\ell_{(3)}^{ij}:=\big(f_{ij}-g_{ij} f^{(3)}\big)^{-1}$ and use it to solve for the vector~$\tilde\Omega^j$ and thus obtain a single equation for the final shock profile~${\cal F}_{oo}$. This gives us a {\it sufficient} condition for vanishing of the characteristic determinant:
\begin{equation}\label{zip}
0=-\frac{3 m^2}{2}+\ell^{\mu}_o \big[{\bar R}_{\mu\nu}{}^{\nu}{}_o+K_{\mu\nu\rho} K^{\nu\rho}{}_o\big] -\frac 12\, \tilde K_i\ell^{ij}_{(3)}
[f\times K\ell]_j\, .
\end{equation}
Absence of superluminal propagation therefore requires, as a {\it necessary} (but {\it not} sufficient) condition that this combination of fields never vanishes.
However it is easy to see that it can: For example, let us focus on flat backgrounds and configurations such that~$K_{ioo}$ is the only non-vanishing contorsion
so that we only need to keep the first and third terms in Eq.~(\ref{zip}) which become~$-3m^2/2+ \ell_{oo} K^j{}_{oo} K_{joo}$. Hence if the normal
$\xi_\mu$ to the characteristic surface is not timelike with respect to the fiducial metric, this quantity is the difference of two positive terms, one of them field dependent, and so clearly can vanish. This confirms the existence of superluminal propagation in the model\footnote{A similar conclusion has also been reached in~\cite{Cham}, who claim the constraint analysis of~\cite{HR} is flawed because it missed extra terms (arising from zeros in the action's Hessian) built from the  time derivative of the metric squared.}.

We now  discuss \emph{acausality}. Up to now, technically we have only established \emph{superluminality}, {\it i.e.}, that gravity excitations can propagate outside of local (metric) light cone. This defect, among other problems, signals that the theory could be \emph{acausal}: it might permit closed timelike curves (CTCs). To see how acausality can arise in mGR, let us consider a special case in which the  fiducial metric is Minkowski (say)  and  the contorsion components $K_{\mu\nu o}=0$ so that ${\cal F}_{oo}=0$. Then we obtain, in an obvious matrix notation, the condition on the remaining shock profiles $\Omega_{ojk}$,
$$
\{f, \Omega\}=0, 
$$
where~$f_{ij}$ is symmetric with respect to the spacelike metric~$g_{ij}$ and can be diagonalized with 
eigenvalues~$(f_1,f_2,f_3)$ . Then, non-vanishing of every pair $(f_1+f_2,f_2+f_3,f_1+f_3)$ is  the necessary and sufficient condition for~$\{f, \Omega\}=0$ to imply~$\Omega=0$. Na\"ively, one might think that the 
eigenvalues of~$f$ must be positive because~$f_{ij}$ seems to be spacelike; however, spacelike-ness with respect to~$f$ and~$g$ will in general \emph{not} coincide. So situations like~$f_1=-f_2$ can occur. 
Consequently, in this setting, it is likely that all spacelike hypersurfaces can be characteristic hypersurfaces, which in turn implies that we could locally embed a closed timelike curve into the spacetime. To summarize, our analysis implies   superluminal propagation and in addition the stronger statement that  (at least) some solutions suffer acausalities\footnote{See also the second entry of~\cite{Gr} for the same conclusion in the model's decoupling limit.}. 

We stress that the acausality that appears here differs from  GR's  CTCs in  two ways:
Firstly, mGR  acausalities arise dynamically and affect asymptotic observers, while in GR dynamical acausalities are difficult to generate without breaking energy conditions or evading protective black hole event horizons (see~\cite{DJtH}). 
Secondly, our acausality is \emph{local}, whereas CTCs in GR are non-local structures: even on CTC solutions,  local, GR,  time evolution is well-defined. Instead, mGR's acausality means that  local  time evolution is not well-defined even in an infinitesimal region.
Therefore, while  GR's acausal solutions are in this sense artificial,  this is \emph{not} the case for mGR's. 
In mGR, as we have shown, causality can be easily violated in the sense that acausal structures can be dynamically formed in local regions: mGR acausality is  far more calamitous than that of GR.

\section{A Toy Model Realization}

Non-linear massive theories face severe consistency problems when made to self-interact or interact with backgrounds, 
so our mGR no-go results come as little surprise given the (second order) findings of~\cite{DW}. 
However, the relation between our first order and that
analysis  is of some interest, especially since the second order superluminality conclusions followed independently of the mass parameter, which indicates that mGR is likely inconsistent, even when employed as an effective theory.
 The latter examined solutions where all field discontinuities 
were of second order and focused accordingly on the leading derivative terms in the second order field equations and constraints. This amounts
to solving the system in an eikonal limit~$g_{\mu\nu}\sim \exp(is \xi . x)\gamma_{\mu\nu}$ with~$s\to \infty$. The result was superluminal 
propagation of the lowest helicity mode for any background or mass term. That is, the ``characteristic matrix'' for second order shocks
found there was not
of maximal rank. On the other hand, our first order shock analysis, which calculates the characteristic matrix in the strict PDE sense---to which one can apply machinery, such as Cauchy--Kowalevski's~(see, for example~\cite{IO}), to deduce evolution of Cauchy data---leads to 
a generically maximal rank\footnote{This first order result, of course, was guaranteed by the correctness of previous ADM-type DoF computations~\cite{HR}.}, but field-dependent, matrix whose zeros as a function of field space lead to superluminal propagation. Although the conclusions are the same, these results might seem contradictory. We therefore introduce
 a simple (but equally pathological) toy model that both explains
how this situation can arise and exhibits both types of superluminalities: Consider a scalar field with action~$S(\varphi)=\int[\frac12\nabla_\mu \varphi \nabla^\mu \varphi+ \frac14(\nabla_\mu \varphi \nabla^\mu \varphi)^2]$ in some non-dynamical background.
The equations of motion can be brought to a simpler, still two-derivative, form by introducing a second, auxiliary, field~$\psi$:
\begin{eqnarray}
\square \varphi + \nabla_\mu (\psi \nabla^\mu \varphi) &=& 0,\nonumber\\[2mm]
\nabla_\mu \varphi \nabla^\mu \varphi-\psi &=& 0\, .\label{toy}
\end{eqnarray}
Indeed, the above system of equations is very similar to the mGR  scalar constraint and leading dynamical equations
of motion. (The fields~$(\psi,\varphi)$ are analogous to~$(g_{oo},g_{ij})$, the first equation being the dynamical one and the second mimicking the scalar constraint.) In particular, in the mGR setting, superluminal behavior of metric components, that happen to be auxiliary in a particular 3+1 
decomposition, is clearly undesirable--even in this simple model, as a second order shock analysis {\it \`a la}~\cite{DW} shows. This demonstrates that the composite operator~$\nabla_\mu \varphi \nabla^\mu \varphi$ is tachyonic and thus unphysical. In eikonal language,
taking~$(\varphi,\psi) =  \exp(is\ \xi . x)\cdot(\Phi, \Psi)$ and~$s$ large, we find~$\Phi +  \Phi\star \Psi=0=\Phi\star \Phi$  (where~$\Phi\star \Psi$ denotes Fourier convolution) so~$\Phi=0$ and~$\Psi$ arbitrary gives superluminal solutions. Of course, second order shocks in~$\psi$ will source third (and possibly higher) order (superluminal) shocks in $\varphi$, a typical feature of models with pathological kinetic terms. To study, on the other hand, the leading second order shocks in $\varphi$, a first order shock analysis of the equations~(\ref{toy}) is needed. 
(In this simple toy case, one can also read off the characteristic determinant from the  original equation of motion~$\big(\square + \nabla^\mu \, [\nabla^\nu \varphi \nabla_\nu \varphi]\,  \nabla_\mu\big) \varphi =0$, and finds~$1-3(\nabla_{\!o\,}\varphi)^2 +(\nabla_i\varphi)^2$, whose zeros again signal superluminal behavior.) In a  first order reformulation we set~$v_\mu=\nabla_\mu\varphi$ and study the system of equations~$(1+\psi)\nabla_\mu v^\mu + v^\mu \nabla_\mu \psi= 0 = v^\mu v_\mu -\psi = v_\mu-\nabla_\mu \psi = \nabla_\mu v_\nu - \nabla_\nu v_\mu$. 
Now, in the same notations as earlier denoting shock profiles by capital letters, we  have~$ \Phi = 0 = V_i$
and thus  the characteristic matrix 
$$
\begin{pmatrix}v_o & 1+\psi\\ 1 & 2v_o\end{pmatrix}\begin{pmatrix}\Psi \\ {\rm V}_o\end{pmatrix} =0\, ,
$$
whose  determinant is (again)~$1+\psi-2v_0^2=1-3(\nabla_{\!o\,}\varphi)^2 +(\nabla_i\varphi)^2$.

An issue that often arises in the context of superluminality is its relation to acausality, since the former may not always imply the 
latter~\cite{Geroch}. Indeed, it has recently  been suggested that for a hyperbolic system of PDE formulated on some spacetime, the causal structure defined by the system's own
evolution (even if superluminal with respect to the background fiducial metric) is the only relevant one~\cite{Bruneton}. This argument does not apply to mGR for two reasons: Firstly, in mGR one of the fields is a dynamical metric, to which matter fields will couple. This field defines local 
light cones and causality--gravity, and light, waves should obey the same caustics, which they manifestly need {\it not} do here. 
Both the first and second order shock analysis demonstrate a failure of causality in this sense. Furthermore, the zeros in the first order characteristic matrix, exhibited in this paper, imply a positivity violation of the kinetic matrix for physical excitations.
Consequently this implies classical instabilities (which have already been found~\cite{MU}), and negative norms in the quantum version of the theory.

\section{Conclusions}

We now summarize our findings for mGR. The presence of tachyonic ``gravitons'', and their deleterious effect on the matter sources with which they unavoidably interact, means the theory could at best be an effective one, within some putative domain of validity. However, the second order analysis in~\cite{DW} shows that there is no such domain, because the tachyons 
were entirely mass- and background- independent.
One might still attempt to argue that, in the first order computation of Section~\ref{sec5} that was based on Cauchy-Kowalevski machinery, superluminal propagation
required special field configurations and acausality was exhibited only in a Minkowski background. 
However, clearly the same mechanism can produce CTCs in more general backgrounds than our simple example's.
Moreover, only the very small graviton (or Vainshtein) mass~\cite{Va} is likely to separate (putative) subsectors free of superluminalities/acausalities from the badly behaved ones,
 so  attempting to save the theory by recourse to effective field theory reasoning seems doomed.
The best hope of avoiding acausality would be to remove the offending fifth DoF in favor of a 4 DoF, de Sitter (or Einstein)-background partially massless model, but this avenue has been exhaustively~\cite{DSW,deRhamPM} excluded. Nor do matters seem any better for the two-tensor bimetric model, according to a recent analysis~\cite{DSW2}. 

Our work emphasizes the importance of the \emph{right kind} of non-linearity for a viable theory of gravity (see also~\cite{OINC}).
For example, even linearized gravity is problematic, but this
is cured by full GR, which emerges through combining the sum of background and spin~2 field excitations into a single, background-independent, dynamical/geometric tensor~\cite{Deser1970}.
The point is that any modified theory of gravity with extra degrees of freedom needs to suppress these new DoF's to recover well-tested GR at the linearized limit, \emph{and} to excite the new DoF's in some regimes (so that one may model (say) dark matter or dark energy with the new DoF's). It is however, not an easy task to excite them \emph{without} attendant problems like ghost modes, superluminality, or acausality. Our analysis thus shows that mGR does \emph{not} seem to give the right kind of non-linearity. 
One might perhaps set one's hope on 
 the last of the remaining, unanalyzed mGR mass term (cubic in the fiducial vierbein), but  we suspect
that it will meet the same fate as the other two choices. A final route to a consistent massive gravity model is to search for some sort
of ``protective'' embedding  (perhaps analogous to that of charged higher spin~\cite{Porrati} and multi-graviton models~\cite{Kiritsis} in string theory). This would entail modifying the Einstein--Hilbert kinetic terms~\cite{HB1}, an inherently dangerous endeavor likely to ruin the constraint structure that mGR inherits from its GR neighbor.
Thus, the philosophically satisfying uniqueness of GR remains solid.

\section*{Acknowledgements}
We thank M. Porrati for catching an important typo. S.D.'s work is supported, in part, by NSF PHY- 1266107 and DOE DE- FG02-164 92ER40701 grants; K.I.'s  by the Taiwan National Science Council under Project No. NSC101-2811-M-002-103 and  Y.C.O.'s by a Taiwan Scholarship from Taiwan's Ministry of Education.

\end{document}